\documentstyle[pra,aps]{revtex}
\begin{document}
\draft
%\twocolumn[\hsize\textwidth\columnwidth\hsize\csname
%@twocolumnfalse\endcsname

\title{Classical quasi-particle dynamics in trapped Bose condensates}

\author{Martin Fliesser}
\address{Fachbereich Physik,
Universit\"at-Gesamthochschule Essen,\\
45117 Essen,Germany}
\author{Andr\'as Csord\'as,} 
\address{Research Group for Statistical Physics of the
Hungarian Academy of Sciences,\\
M\'uzeum krt. 6--8, H-1088 Budapest,
Hungary}
\author{Robert Graham,}
\address{Fachbereich Physik,
Universit\"at-Gesamthochschule Essen,\\
45117 Essen,Germany}
\author{P\'eter Sz\'epfalusy}
\address{Institute for Solid State
Physics, E\"otv\"os University,\\
M\'uzeum krt. 6--8, H-1088 Budapest,
Hungary, and\\
Research Institute
for Solid State Physics, P.O. Box 49, H--1525 Budapest, Hungary,}

\date{\today}
\maketitle

\begin{abstract}
The dynamics of quasi-particles in repulsive Bose condensates in a
harmonic trap is studied in the classical limit. In isotropic traps
the classical motion is integrable and separable in spherical coordinates.
In anisotropic traps the classical dynamics is found, in general, to be
nonintegrable. For quasi-particle energies  $E$ much smaller than the
chemical potential $\mu$, besides the conserved quasi-particle energy,
we identify two additional nearly conserved phase-space functions. These
render the dynamics inside the condensate (collective dynamics) integrable
asymptotically for $E/\mu\to0$. However, there coexists at the same energy
a dynamics confined to the surface of the condensate, which
is governed by a classical Hartree-Fock Hamiltonian. We find that also this
dynamics becomes integrable for $E/\mu\to0$ because of the appearance of an adiabatic invariant. For $E/\mu$ of order 1
a large portion of the phase-space supports chaotic motion, both, for the Bogoliubov Hamiltonian and its Hartree-Fock approximant. To exemplify
this we exhibit Poincar\'e surface of sections for harmonic traps with the
cylindrical symmetry and anisotropy found in TOP traps. For $E/\mu\gg1$ the
dynamics is again governed by the Hartree-Fock Hamiltonian. In the
case with cylindrical symmetry it becomes quasi-integrable because the
remaining small chaotic components in phase space are tightly confined
by tori.  
\end{abstract}


%\vskip2pc]
%\narrowtext

%%%%%%%%%%%%%%%%%%%%%%%%%%%%%%%%%%%%%%%%%%%%%%%%%%%%%
\section{Introduction}
%%%%%%%%%%%%%%%%%%%%%%%%%%%%%%%%%%%%%%%%%%%%%%%%%%%%%

The experimental realization of Bose-condensates of atoms harmonically
bound in magnetic traps 
\cite{anderson,bradley,davis}
call for a space-dependent version of Bogoliubov's
theory, or some modification thereof. Such a theory proceeds by splitting
the field operator $\hat{\psi}(\bbox{x})$ and its adjoint in a $C$-number
part $\psi_0(\bbox{x})$ and a residual operator $\hat{\varphi}(\bbox{x})$,
\begin{equation}
\label{eq:1}
 \hat{\psi}(\bbox{x})=\psi_0(\bbox{x})+\hat{\varphi}(\bbox{x})
\end{equation}
and an accompanying decomposition of the Hamiltonian in terms of 0, 1, 2,
3, 4 order in $\hat{\varphi}$, $\hat{\varphi}^+$. The term of 1 order in
$\hat{\varphi}$, $\hat{\varphi}^+$ is made to vanish by choosing 
$\psi_0(\bbox{x})$ to satisfy the time-independent Gross-Pitaevskii
equation \cite{gross}, 
which at low temperatures, takes the form
\begin{equation}
\label{eq:2}
-\frac{\hbar^2}{2m}\nabla^2\psi_0(\bbox{x})+(U(\bbox{x})-\mu)
 \psi_0(\bbox{x})+V_0|\psi_0(\bbox{x})|^2\psi_0(\bbox{x})=0\,,
\end{equation}
with the normalization $\int|\psi_0|^2d^3x=N_0\,$. Here
\begin{equation}
\label{eq:3}
 U(\bbox{x})=\frac{m}{2}(\omega^2_x x^2+\omega_y^2 y^2+\omega_z^2 z^2)
\end{equation}
is the generally anisotropic harmonic trap potential,
\begin{equation}
\label{eq:4}
 V_0=\frac{4\pi\hbar^2a}{m}
\end{equation}
is the strength of the pseudo-potential replacing the true two-particle
potential at low energies, with the $s$-wave scattering length $a$, which
is here assumed to be positive.

For $(N_0a/d_0)\gg1$, where $d_0=\sqrt{\hbar/m\bar{\omega}}$,
$\bar{\omega}=(\omega_x\omega_y\omega_z)^{1/3}$, the solution to the
Gross-Pitaevskii equation can be determined in the Thomas-Fermi
approximation \cite{baym} 
by neglecting the kinetic-energy term
\begin{equation}
\label{eq:5}
 |\psi_0|^2=\frac{\mu-U(\bbox{x})}{V_0}\Theta(\mu-U(\bbox{x}))\,.
\end{equation}
In the following we shall choose $\psi_0$ as real and positive. The chemical potential is determined from the normalization. The
next step is the diagonalization of that part of $H$, which is a quadratic
form in $\hat{\varphi}$, $\hat{\varphi}^+$, by a Bogoliubov transformation
to quasi-particles
\begin{equation}
\label{eq:6}
 \hat{\varphi}(\bbox{x})=\sum_j
  \left(U_j(\bbox{x})\hat{\alpha}_j-V_j^*(\bbox{x})
   \hat{\alpha}_j^+\right)
\end{equation}
with
\begin{equation}
\label{eq:7}
 \int d^3x\left(|U_j(\bbox{x})|^2-|V_j(\bbox{x})|^2\right)=1
\end{equation}
and
\begin{eqnarray}
\label{eq:8}
 [\alpha_j,\alpha_{j'}] &=& 0=[\alpha_j^+,\alpha_{j'}^+]\nonumber\\
 ~[\alpha_j,\alpha_{j'}^+] &=& \delta_{jj'}\,.
\end{eqnarray}
The second-order part of $H$ is diagonalized by this transformation, if
$U_j(\bbox{x})$ and $V_j(\bbox{x})$ satisfy the Bogoliubov equations
\cite{fetter}
\begin{equation}
\label{eq:9}
 \left(\begin{array}{c@{\quad}c}
 \hat{H}_{\mbox{\tiny HF}} & -K(\bbox{x})\\
 -K(\bbox{x}) & \hat{H}_{\mbox{\tiny HF}}\end{array}\right)
 \left(\begin{array}{c}
  U_j\\  V_j\end{array}\right)
  =E_j\left(\begin{array}{c}
  U_j\\ -V_j\end{array}\right)
\end{equation}
with the Hartree-Fock Hamiltonian
\begin{equation}
\label{eq:10}
 \hat{H}_{\mbox{\tiny HF}}=-\frac{\hbar^2}{2m}\nabla^2+U(\bbox{x})
  -\mu+2V_0|\psi_0(\bbox{x})|^2
\end{equation}
and the coupling term
\begin{equation}
\label{eq:11}
 K(\bbox{x})=V_0|\psi_0(\bbox{x})|^2
\end{equation}
between the two components $U_j(\bbox{x})$, $V_j(\bbox{x})$ of a
quasi-particle wave-function. Because of the different signs of the
$U_j$, $V_j$ components on the right-hand side, they play the role
of particle and anti-particle components of the complete wave-function.
As the equations are symmetric under the particle-antiparticle
transformation $E_j\to -E_j$, $U_j\to V_j^*$, $V_j\to U_j^*$ we may
define $E_j$ to be non-negative without restriction of generality.
Various numerical \cite{numerical} 
and approximate analytical \cite{analytical} 
treatments of these equations are available in the literature.

In the present paper we wish to study the classical limit of the
center-of-mass motion of the quasi-particles. In order to discuss the
dynamics rather than the eigenstates of the quasi-particles, it is
useful to introduce time-dependent wave functions via
\begin{equation}
\label{eq:12}
 \left(\begin{array}{c}
  U(t)\\V(t)\end{array}\right)=\sum_jc_j
 \left(\begin{array}{c}
  U_j\\V_j\end{array}\right)e^{-iE_jt/\hbar}
\end{equation} 
with arbitrary coefficients $c_j$. They satisfy the time-dependent
Schr\"odinger equation
\begin{equation}
\label{eq:13}
 i\hbar\frac{\partial}{\partial t}
  \left(\begin{array}{c}
   U(t)\\-V(t)\end{array}\right)=
  \left(\begin{array}{c@{\quad}c}
   \hat{H}_{\mbox{\tiny HF}} & -K\\
    -K & \hat{H}_{\mbox{\tiny HF}}\end{array}\right)
  \left(\begin{array}{c}
   U(t)\\V(t)\end{array}\right)\,. 
\end{equation}
For large energies $E_j$, $E_j\gg\mu$, the classical motion can be
interpreted as the center of mass-motion of quasi-particle
wave packets. For small energies $E_j$, $E_j\ll\mu$, such a straightforward
physical interpretation of the classical quasi-particle dynamics is no
longer possible. However, even in this regime, there is still a close
mathematical relation between the classical and the quantum dynamics,
as the classical trajectories are the characteristics of the quantum
mechanical wave-equation. This is made explicit by the derivation of the
classical dynamics as a limit of the Schr\"odinger equation via the
Hamilton-Jacobi equation. The Hamilton-Jacobi equation corresponding to
eq.~(\ref{eq:13}) is obtained by the asymptotic ansatz for $\hbar\to0$
\begin{equation}
\label{eq:14}
 \left(\begin{array}{c}
  U(\bbox{x},t)\\ V(\bbox{x},t)\end{array}\right)\simeq
 \left(\begin{array}{ccc}
  a_0(\bbox{x},t) & + & 0(\hbar)\\
  b_0(\bbox{x},t) & + & 0(\hbar)\end{array}\right)
 e^{iS(\bbox{x},t)/\hbar}
\end{equation}
with $\int d^3x(|a_0(\bbox{x},t)|^2-|b_0(\bbox{x},t)|^2)=1$.
It reduces eq.~(\ref{eq:13}) to the form, to zeroth order in $\hbar$,
\begin{equation}
\label{eq:15}
\left(\begin{array}{cc}
 \epsilon_{\mbox{\tiny HF}}+\frac{\partial S}{\partial t} & -K\\
  -K & \epsilon_{\mbox{\tiny HF}}-\frac{\partial S}{\partial t}
   \end{array}\right)
\left(\begin{array}{c}
 a_0\\ b_0\end{array}\right)=0\,.
\end{equation}
Here
\begin{equation}
\label{eq:16}
 \epsilon_{\mbox{\tiny HF}}=\frac{p^2}{2m}+U(\bbox{x})-\mu+2V_0
  |\psi_0(\bbox{x})|^2\,.
\end{equation}
We may restrict to $-E=\frac{\partial S}{\partial t}\le0$ in accordance
with our restriction on $E$. To first order in $\hbar$ we obtain
\begin{equation}
\label{eq:17}
\frac{\partial}{\partial t}
 \left(\begin{array}{l}
  a_0\\ -b_0\end{array}\right)+\frac{1}{2m}\bbox{\nabla \cdot}
   \left((\bbox{\nabla}S)
  \left(\begin{array}{l}
  a_0 \\ b_0\end{array}\right)\right)+\frac{1}{2m}\bbox{\nabla}S 
   \bbox{\cdot \nabla}
   \left(\begin{array}{l}
  a_0 \\ b_0\end{array}\right)=\frac{i}{\hbar}
  \left(\begin{array}{cc}
   \epsilon_{\mbox{\tiny HF}}+\frac{\partial S}{\partial t} & -K\\
   -K & \epsilon_{\mbox{\tiny HF}}-\frac{\partial S}{\partial t}
    \end{array}\right)
   \left(\begin{array}{l}
  a_1 \\ b_1\end{array}\right)\,.
\end{equation}
Here ${a_1 \choose b_1}$ are the $0(\hbar)$-components of the amplitudes
in (\ref{eq:14}). These will exist, and the expansion will be well-defined,
only if the left-hand side of (\ref{eq:17}) is orthogonal on the kernel
${a_0 \choose b_0}$ of the matrix in (\ref{eq:15}), which also appears on the
right-hand side of (\ref{eq:17}). This condition gives rise to the
conservation law
\begin{equation}
\label{eq:17a}
 \frac{\partial}{\partial t}(|a_0|^2-|b_0|^2)+\frac{1}{2m}\bbox{\nabla \cdot}
  ((|a_0|^2+|b_0|^2)\bbox{\nabla}S)=0
\end{equation}
which ensures that the normalization condition
\[
 \int d^3x(|a_0|^2-|b_0|^2)=1
\]
is consistent with the classical dynamics and represents
the classical limit of the continuity equation following from
(\ref{eq:17a}) \cite{csordas}. The zeroth order equation has
a nontrivial solution only if the determinant condition
\begin{equation}
\label{eq:18}
 \left(\frac{\partial S}{\partial t}\right)^2=\epsilon_{\mbox{\tiny HF}}^2
  -K^2
\end{equation}
is satisfied, which, observing our sign convention for
$\partial S/\partial t$, gives the time-dependent Hamilton-Jacobi equation
\begin{equation}
\label{eq:19}
 \frac{\partial S(\bbox{x},t)}{\partial t}+H
  \left(\bbox{x},\frac{\partial S}{\partial\bbox{x}}\right)=0
\end{equation}
with the classical Hamiltonian
\begin{equation}
\label{eq:20}
 H(\bbox{x},\bbox{p})=
  \sqrt{\epsilon_{\mbox{\tiny HF}}^2(\bbox{x},\bbox{p})-K(\bbox{x})^2}\,.
\end{equation}
The time-independent Hamilton Jacobi equation results from the separation
\begin{equation}
\label{eq:21}
 S(\bbox{x},t)=S(\bbox{x})-Et
\end{equation}
and reads
\begin{equation}
\label{eq:22}
 H(\bbox{x},\frac{\partial S}{\partial\bbox{x}})=E\,.
\end{equation}
If we can neglect the $0(\hbar)$-corrections $a_1$,$ b_1$ in (\ref{eq:17})
we obtain from the
first order equation separate conservation laws for the quasi-particle
and anti-quasi-particle densities
\begin{eqnarray}
\label{eq:23}
\frac{\partial}{\partial t}|a_0(\bbox{x},t)|^2+\frac{1}{2m}\bbox{\nabla \cdot}
 \left(|a_0(\bbox{x},t)|^2\bbox{\nabla}S\right)&=& 0\nonumber\\
\frac{\partial}{\partial t}|b_0(\bbox{x},t)|^2-\frac{1}{2m}\bbox{\nabla \cdot}
 \left(|b_0(\bbox{x},t)|^2\bbox{\nabla}S\right)&=& 0 \,.
\end{eqnarray}
The classical anti-particle and particle dynamics are therefore just the
time-reversed of each other, and the densities of both components are
separately conserved.

In the following sections we analyze the classical dynamics described by the
Hamiltonian (\ref{eq:20}).

%%%%%%%%%%%%%%%%%%%%%%%%%%%%%%%%%%%%%%%%%%%
\section{Classical quasi-particle dynamics}
%%%%%%%%%%%%%%%%%%%%%%%%%%%%%%%%%%%%%%%%%%%

For the case of isotropic harmonic traps angular momentum is conserved
and the quasi-particle dynamics is integrable and separable in spherical
coordinates. This case is discussed in \cite{csordas}, 
where it is made the
basis of a semiclassical quantization procedure. Therefore, in the
following we concentrate on the analysis of the case of anisotropic
harmonic traps in the limit where the Thomas-Fermi approximation applies.
In the present section we shall assume cylindrical symmetry of the trap
\begin{equation}
\label{eq:24}
U(\bbox{x})=\frac{m\omega_0^2}{2}(x^2+y^2)+\frac{m\omega_z^2}{2}z^2
\end{equation} 
In the experiment \cite{anderson} 
$\omega_z > \omega_0 $, namely,
\ $(\omega_z/ \omega_0)^2 \approx 8$.
As the parameter denoting the anisotropy of the potential we introduce
$\epsilon$ by
$\epsilon^2=1-(\omega_0/\omega_z)^2\,$, which is the numerical excentricity
of the Thomas-Fermi surface $\mu=U(\bbox{x})$, a rotational 
symmetric ellipsoid. 
This two-dimensional
surface is the boundary of the condensate.

Our problem has a characteristic energy, namely the chemical
potential. Thus, the second relevant parameter of the classical
motion is the ratio $E/\mu$. We note that measuring the energy in
units of $\mu$, coordinates, momenta and time in units of
\begin{equation}
   r_0=\sqrt{\frac{2\mu}{m\omega_0^2 }} 
   \quad ,\quad 
   p_0=\sqrt{2m\mu} 
   \quad ,\quad 
   t_0= \omega_0^{-1}
      \label{scale.param}
\end{equation}
respectively, the dimensionless Hamiltonian can be put in
a form, which depends only on the anisotropy parameter $\epsilon$.
This shows that condensates with the same anisotropy
but with different chemical potential behave similarly
in the classical description, if the physical quantities 
are scaled appropriately.

In the isotropic case $\epsilon=0$ the classical dynamics 
are completely integrable.
As three independent constants of motion we can choose the energy $E\,$, 
the modulus of the angular momentum 
and its $z$-component $L_z$.
As we keep rotational symmetry around the z-axis 
in the anisotropic case $\epsilon \ne 0$ 
the $L_z$ and of course the energy
are still conserved quantities,
whereas the total angular momentum considered here
is no longer a constant of motion.
Thus, in the following we shall investigate the classical behaviour 
of this three degrees of freedom system depending 
on the two constants of motion $E$ and $L_z$,
and we address the question wether the dynamics are integrable or chaotic.

Let us introduce the usual cylindrical coordinates 
$\rho = \sqrt{x^2+y^2}$, z
and $\phi$. Because of the rotational symmetry around of the
$z$-axes the angle
$\phi$ is a cyclic variable. In cylindrical
variables the Hamiltonian has merely two degrees of freedom
$\rho$ and $z$,
$L_z$ just enters as a parameter.
Certain conditions have to be satisfied
as can be seen from the Hamiltonian in the region outside
the condensate. For $E>\mu$ the condition $E+\mu > \omega_0 L_z$
has to be guaranteed,
for $E < \mu$ we must have
$E>(\omega_0 L_z)^2/4\mu\,$.

%%%%%%%%%%%%%%%%%%%%%%%%%%%%%%%%%%%%%%%%%%%%%%%%%%%%%%%%
% Numerical stuff for the behavior of the motion 
% with the full Hamiltonian
%%%%%%%%%%%%%%%%%%%%%%%%%%%%%%%%%%%%%%%%%%%%%%%%%%%%%%%%
The dynamics of this two dimensional system we can 
visualize by Poincar\'e cuts,
see Fig.\ref{poinc.cuts}.
For different energies we observe different dynamical behaviour.
For $E>(\omega_0 L_z)^2/4\mu>\mu$
two different kinds of trajectories can occur typically.
If the repulsive effective potential in $\rho$-direction
due to the angular momentum $L_z$ is strong enough,
the particle cannot enter the condensate and
is only moving in the harmonic potential of the external trap.
The motion in an anisotropic harmonic potential is completely integrable,
as a third constant of motion we can choose the energy 
in the z-degree of freedom
$ E_z = p_z^2/2m + m \omega_{z} z^2/2\,.$
These trajectories, which are not perturbed by the condensate,
can be seen as the integrable tori
around the fixed point of the Poincar\'e map
in the centre of Fig.\ref{poinc.cuts}a,
which is the periodic orbit moving only in $z$ and $\phi$ directions.
If the particle enters the condensate,
$E_z$ is no more a conserved quantity.
Nevertheless for energies large compared to the chemical potential
also those trajectories
are still quite similar to unperturbed motion.
Typically the trajectories are confined to thin stochastic layers
separated by each other by integrable tori.
No Arnold diffusion occurs,
as usually for a system of two, not of three degrees of freedom.
At high energies the system behaves quasi integrable. 
The influence of the condensate can be taken as a small perturbation
to the motion in the external potential.

For energies in the range $10>E/\mu>0.1$ (Fig.\ref{poinc.cuts}b)
we typically observe a mixed phase space.
The fixed point is now inside the condensate,
but does not loose its stability.
The detailed structure depends on the parameters chosen.
Already for small anisotropy ($\epsilon^2=0.2$)
a relevant part of phase space can be chaotic.
This shows that for energies of the order of the chemical potential
the isotropic case with its integrable dynamics
is an exceptional rather than a typical situation.
If $E<\mu$ all trajectories move inside and outside the condensate.

For energies small compared to the chemical potential $E<0.1\mu$
(Fig.\ref{poinc.cuts}c) the chaotic part
of phase space decreases again and
is restricted to a thin layer 
separating and surrounding two regular islands,
corresponding to two stable fixed points separated by an unstable one.
Most orbits seem to lie on integrable tori.
This suggests that the system has an integrable regime
in the limit of small energies.

%%%%%%%%%%%%%%%%%%%%%%%%%%%%%%%%%%%%%%%%%%%
% The hydrodynamical region
%%%%%%%%%%%%%%%%%%%%%%%%%%%%%%%%%%%%%%%%%%%
This limit corresponds to the hydrodynamical regime \cite{stringari} 
investigated in several contexts. In a bulk case, when there
is no external potential $U(\bbox{x})$ the lowest lying excitations
are phonons with linear wave-number dependence. 
%Some more sentences about this limit.

Numerically we have found that tending with the energy to zero,
keeping $\mu$ fixed
the range of the classical motion outside the condensate for trajectories starting inside is 
getting smaller and smaller and  in the 
limit the motion is confined to the region inside of the
Thomas-Fermi surface. 
Starting trajectories from the same point inside the condensate
under the same direction and changing 
only the modulus of Cartesian-momentum we have found that 
they differ from each other only
in a thin region near the boundary whose width scales 
with the energy. Lowering the modulus of the
initial momentum
to zero they tend to a well-defined limiting trajectory.
This can be clearly seen in Fig.\ref{limit.trajects}.
In the isotropic case this is the limit $E/\mu,\omega_0L/\mu \to 0\,$,
keeping the ratio $L/E$ fixed. In the following section this 
'hydrodynamic regime' will be studied in detail for anisotropic 
traps. However, 
it will turn out that in traps there exists a second low-energy 
regime, which for isotropic traps is defined by $E/\mu \to 0$ 
with $E-(\omega_0 L)^2/4\mu \ll E$, where the quasi-particles 
are single-particle like excitations confined to a narrow 
layer around the surface of the condensate. This low-energy 
Hartree-Fock regime will be discussed in detail in 
section IV, also for anisotropic traps, together with
the usual high-energy ($E\gg\mu$) Hartree-Fock regime.
%%%%%%%%%%%%%%%%%%%%%%%%%%%%%%%%%%%%%%%%%%%%%%%%%%%%%%%%%%%%

\section{Quasi-particle dynamics in the hydrodynamic regime}
%%%%%%%%%%%%%%%%%%%%%%%%%%%%%%%%%%%%%%%%%%%%%%%%%%%%%%%%%%%%

\subsection{Hydrodynamic Hamiltonian}
If there exist limiting trajectories for different 
initial conditions there should exist limiting dynamics 
described by some limiting Hamilton-function. 
Inside the condensate the Bogoliubov Hamiltonian can be written as
\begin{equation}
   H(\bbox{p},\bbox{x}) 
  =\sqrt{\epsilon_{kin}(\bbox{p})
        (\epsilon_{kin}(\bbox{p}) + 2 K(\bbox{x}))}
   \,, \label{bogol.ham.anders}
\end{equation}
where  $\epsilon_{kin}(\bbox{p})=\bbox{p}^2/2m$.
For small energies $K(\bbox{x})$ is much bigger than $\epsilon_{kin}$
everywhere except in a small region near the boundary.
This suggests that the approximant of the Hamilton 
function (\ref{bogol.ham.anders}) can be obtained
by neglecting the kinetic energy square
\begin{equation}
   H_{hyd}(\bbox{p},\bbox{x}) 
  =\sqrt{2\,\epsilon_{kin}(\bbox{p})\,K(\bbox{x})}
   \label{hydro.ham}
\end{equation}
for describing the motion in the hydrodynamical regime.
This approximate Hamiltonian is in accordance with the
bulk case, when $K(\bbox{x})=\mu$ should be taken in (\ref{hydro.ham}) 
in order to obtain
the linear phonon spectra from the Bogoliubov dispersion relation.

This Hamiltonian is meaningful only inside the condensate and
only near the boundary of the condensate
the full Hamiltonian (\ref{bogol.ham.anders})
differs from this approximate one.
On the Thomas-Fermi surface the full Hamilton function 
gives definite values for the Cartesian
momenta,
whereas according to $H_{hyd}$ they become infinite.
Following the trajectories of $H_{hyd}$ in the isotropic case
the angular momentum conservation requires that the tangential
component of the momentum remains finite even though the absolute
value of the momentum diverges like $K^{-1/2}$. Therefore
each trajectory hits the boundary orthogonally and is reflected back
orthogonally without change in the tagential component of the momentum.
As this local rule is independent of the global symmetry of the
trap potential it must hold also in the anisotropic case. 

The Hamiltonian (\ref{hydro.ham}) has some further unusual
features.
The first observation is that it is  
not of the usual form but is a homogeneous first order function
of the momenta. The strong consequence is that with the same initial value
$\bbox{x}(t=0)$ and with the same direction of the initial momenta
the orbit $\bbox{x}(t)$ is the same independently of the energy. 
Secondly, a constraint  follows from the canonical equations of motion,
namely
\begin{equation}
   m\bbox{\dot x \cdot \dot x}
   =  \mu - U(\bbox{x})
   \,, \label{constraint}
\end{equation}
relating the velocities and the coordinates. Thus one cannot
choose the initial point and the velocity independently.
Furthermore, due to this constraint one cannot express 
the three velocities in terms of the momenta,
i.e., one cannot do the inverse Legendre transformation in the usual way
to derive the Lagrangian.
From (\ref{constraint}) it is clearly seen that
despite of the divergence of momenta on the boundary of the condensate
the velocities even tend to zero here.

%%%%%%%%%%%%%%%%%%%%%%%%%%%
\subsection{Isotropic case}

In the isotropic trap case ($\omega_0=\omega_x=\omega_y=\omega_z$)
the Poisson-bracket
of $H_{hyd}(\bbox{p},\bbox{x})$  (See Eq.(\ref{hydro.ham})) 
and the angular momentum vector $\bbox{L}$ is zero, which means that
any components of $\bbox{L}$ is a conserved quantity.
Let us choose our coordinate system in such a way that the $z$ axis
is parallel with $\bbox{L}$. In such a frame $L_x=L_y=0$,
which shows that the motion in the phase space
stays on the hypersurface $z=p_z=0$. By this choice of the
coordinates one can eliminate one degrees of freedom from 
the Hamiltonian (\ref{hydro.ham}), which has then the form
\begin{equation}
H_{hyd}(\bbox{p},\bbox{x})=\sqrt{{\mu \over m}
\bigl(p_x^2+p_y^2\bigr)\bigl(1-{x^2+y^2 \over r_{TF}^2} \bigr)},
\label{iso.ham}
\end{equation}
where $r_{TF}$ denotes the radial size of the condensate, 
the Thomas Fermi radius 
$r_{TF}=\sqrt{2\mu \over m \omega_0^2}$.
Let us now consider the transformation
\begin{eqnarray}
x&=& {r_{TF} \over (I_1+I_2)}\bigl( I_1 \cos{\phi_1}+ I_2\cos{\phi_2}\bigr),
\nonumber\\
y&=& {r_{TF} \over (I_1+I_2)}\bigl( I_1 \sin{\phi_1}- I_2\sin{\phi_2}\bigr),
\nonumber\\
p_x&=& -{(I_1+I_2) \over r_{TF}\bigl(1-\cos{(\phi_1+\phi_2)} \bigr)}
(\sin{\phi_1}+\sin{\phi_2}),
\nonumber\\
p_y&=& {(I_1+I_2) \over r_{TF}\bigl(1-\cos{(\phi_1+\phi_2)} \bigr)}
(\cos{\phi_1}-\cos{\phi_2}),
\label{can.transf}
\end{eqnarray}
with positive $I_1$ and $I_2$. It is straightforward to check
that the Poisson-brackets between $I_1,I_2,\phi_1,\phi_2$ are canonical,
thus, the transformation (\ref{can.transf}) is a canonical transformation.
Inserting (\ref{can.transf}) into (\ref{iso.ham}) one gets
\begin{equation}
E=H(I_1,I_2,\phi_1,\phi_2)=\omega_0 \sqrt{2 I_1 I_2},
\label{iso.ham1}
\end{equation}
i.e., $I_1$ and $I_2$ are the action and $\phi_1$ and $\phi_2$ the
angle coordinates of the Hamiltonian (\ref{iso.ham}). Similarly
to the harmonic oscillator case this Hamiltonian is a homogeneous
first order function of the action coordinates.

The Hamilton equations in the new coordinates are
\begin{eqnarray}
\dot{I}_1&=&0 \qquad 
\dot{\phi}_1={\omega_0\over\sqrt{2}}\sqrt{I_2\over I_1} =\Omega_1,
\nonumber \\
\dot{I}_2&=&0  \qquad 
\dot{\phi}_2={\omega_0\over\sqrt{2}}\sqrt{I_1\over I_2} =\Omega_2.
\label{can.eq}
\end{eqnarray}
Using the above transformation, it is easy to show that 
the angular mumentum is
\begin{equation}
L_z = x p_y - y p_x= I_1 -I_2.
\end{equation}
A nice geometrical meaning for $\bbox{x}(t)$ can be given.
Let us consider a circle of radius $b$, in which a smaller
circle of radius $a$ rolls. The motion of a point on the perimeter of the
smaller circle in cartesian coordinates is described by
the equations
\begin{eqnarray}
x&=&(b-a)\cos{\phi_1}+a\cos{\phi_2}, \nonumber \\
y&=&(b-a)\sin{\phi_1}-a\sin{\phi_2},
\label{hypocikloid}
\end{eqnarray}
where $\phi_1$ and $\phi_2$ are linear functions of the time,
see Fig.\ref{pic.cycloid}.
% ($\phi_1$ describes the angle between the x-axes and the line
% pointing towards to the center of mass anticlockwise, while $\phi_2$
% is the angle between the  $x$ direction and the line between the center
% of mass and the system point on the perimeter, clockwise). 
Due to the
perfect rolling condition the angular velocities are not independent:
\begin{equation}
0=(b-a)\dot{\phi_1}-a\dot{\phi_2}.
\label{const1}
\end{equation}
Comparing the parametric form of the hypocycloid (\ref{hypocikloid})
with (\ref{can.transf}) it is obvious that $\bbox{x}(t)$ 
fulfills (\ref{hypocikloid}) and the constraint (\ref{const1}),
if $b=r_{TF}$, $a=r_{TF}I_2/(I_1+I_2)$,
and if $\dot{\phi}_1$,  $\dot{\phi_2}$ are chosen as in (\ref{can.eq}).

The radial distance from the origin can be expressed by
\begin{equation}
r=\sqrt{x^2+y^2}={r_{TF}\over I_1+I_2}
\sqrt{I_1^2+I_2^2+2 I_1 I_2 \cos{(\phi_1+\phi_2)}}.
\label{rad.dist}
\end{equation}
It is obvious that it is periodic
in $(\phi_1+\phi_2)$, its period can be calculated from
$2 \pi=(\Omega_1+\Omega_2)T_r$, which yields
\begin{equation}
T_r={2\pi \over \omega_0}{E \over \sqrt{2E^2+(L_z \omega_0)^2}}.
\end{equation}

The Hamiltonian (\ref{iso.ham}) can be written in polar
coordinates $r$, $\phi$ as well. $\phi$ is a cyclic variable,
its conjugate momentum $I_{\phi}=|L_z|$ is a conserved quantity.
However, the momentum $p_r=(xp_x+yp_y)/r$ conjugated to $r$ is not
conserved.
To express the Hamiltonian (\ref{iso.ham}) in the action variables 
$I_r$, and $I_\phi$ let us use the fact that $2 \pi I_r=\oint p_r\, dr$
and that during one period of the radial motion
$\phi_1+\phi_2$ changes by $2 \pi$.
Using the above formulas one gets
\begin{eqnarray}
I_r&=&\min{(I_1,I_2)}, \nonumber \\
I_\phi&=&| I_1 -I_2 | , 
\end{eqnarray}
which leads by (\ref{iso.ham1}) to
\begin{equation}
E=H_{hyd}(I_r,I_\phi)=\omega_0\sqrt{2\bigl(I_r+I_\phi\bigr)I_r}.
\label{iso.ham2}
\end{equation}

If one quantizes semiclassically the Hamiltonian 
(\ref{iso.ham2}) one should
take into account that in the radial direction there are two
turning points, thus, $I_r$ should be replaced by $\hbar(n+1/2)$,
and by the usual procedure for spherically symmetric problems
$I_\phi$ by $\hbar(l+1/2)$ ($l$ and $n$ are non-negative integers). 
The semiclassical quantization
leads by the above
replacement rules to
\begin{equation}
E_{n,l}=\hbar \omega_0 \sqrt{2n^2+2nl+3n+l+1},
\end{equation}
which is almost that of the result of Stringari \cite{stringari},
except the $1$ under the square-root, and agrees with that of
the more elaborate semiclassical quantization in the hydrodynamical
limit \cite{csordas}.

%%%%%%%%%%%%%%%%%%%%%%%%%%%%%%%%%%%%%%%%%%%%%%%%%%%%%
\subsection{Anisotropic case with cylindrical symmetry}

The case of a trap with axial or cylindrical symmetry is the
experimentally most relevant one.
In Poincar\'e cuts of the full dynamics we have seen regular behaviour
for small energies. Therefore one can expect that the classical 
motion given by the approximate Hamiltonian is fully integrable.
To show this let us introduce new coordinates, namely
the cylindrical elliptical coordinates $\xi,\,\eta$ given by
\begin{equation}
   \rho = \sigma \sqrt{(\xi^2+1)(1-\eta^2)}
   \quad ,\quad 
   z = \sigma \xi \eta 
   \quad
   \label{ellipt.coords}
\end{equation}
which are orthogonal coordinates. 
Surfaces of constant $\xi$ are confocal ellipsoids with foci 
at a distance $\sigma$ in $\rho$ direction,
surfaces of constant $\eta$ are confocal hyperboloids 
with the same foci.
For $\sigma$, the parameter of the transformation, we take the foci 
of the Thomas-Fermi ellipsoid, $\sigma=\epsilon (2\mu/m \omega_0^2)^{1/2}$ 
for $\omega_z>\omega_0$.
For $\omega_0>\omega_z$ one has to change 
$\xi^2+1$ to $\xi^2-1$
and take 
$\sigma=\epsilon (2\mu/m \omega_z^2)^{1/2}$. In the following we
consider only the first case (\ref{ellipt.coords}), in the second case
the analysis proceeds similarly.
$\xi$ can take any value in the range $[0,(1/\epsilon^2-1)^{1/2}]$.
The limiting case $\xi=(1/\epsilon^2-1)^{1/2}$ 
describes the Thomas-Fermi ellipsoid.
$\eta$ can be in the range $[-1,1]$.
Making the point transformation from cylindrical to
cylindrical elliptical coordinates the momenta transform as
\begin{eqnarray}
   p_{\rho} &=&\frac{1}{\sigma}\frac{1}{\xi^2+\eta^2}
             \sqrt{(\xi^2+1)(1-\eta^2)}
            (\xi p_{\xi}-\eta p_{\eta})
   \quad ,\nonumber\\
   p_{z} &=&\frac{1}{\sigma}\frac{1}{\xi^2+\eta^2}
            ((\xi^2+1)\eta p_{\xi}+(1-\eta^2)\xi p_{\eta})
   \label{ellipt.moments}
   \quad.
\end{eqnarray}
The Hamiltonian (\ref{hydro.ham})
in cylindrical elliptical coordinates is
\begin{eqnarray}
  H_{hyd}^2 =
     {\omega_z^2 \over 2 \epsilon^2}
  {(1-\epsilon^2(\xi^2+1))(1-\epsilon^2(1-\eta^2))\over \xi^2+\eta^2 }
  && \nonumber\\
  \times \Bigl((\xi^2+1)p_{\xi}^2 + (1-\eta^2) p_{\eta}^2
       +(\frac{1}{1-\eta^2}-\frac{1}{\xi^2+1})p_{\phi}^2 \Bigr).
   \label{hydro.ham.ellip}
\end{eqnarray}
Taking the energy $E$ and $p_{\phi}=L_{z}$ as constants one can write down 
the Hamilton-Jacobi equation for $\xi$ and $\eta$,
which is separable in these coordinates.
Thus the problem is fully integrable.
Introducing a separation constant $B>0$
the two separated Hamilton-Jacobi equations are
\begin{eqnarray}
   (\xi^2+1)\left(\frac{d S_{\xi}}{d \xi}\right)^2 
  -\frac{L_{z}^2}{\xi^2+1}
  -{2 E^2 \over \omega_z^2}{1 \over 1-\epsilon^2 (\xi^2+1)}
 &=&-B
   \nonumber\\
   (1-\eta^2)\left(\frac{d S_{\eta}}{d \eta}\right)^2 
  +\frac{L_{z}^2}{1-\eta^2}
  +{2 E^2 \over \omega_z^2}{1 \over 1-\epsilon^2(1-\eta^2)}
 &=& B \, .
   \label{hamjacobi.ellipt}
\end{eqnarray}
% We see that the dynamical behaviour only depends on the ratios
% $L_z/E$ and $B/E^2\,$.
Combining these two  equations one gets for the separation constant $B$
the phase-space function 
\begin{eqnarray}
   B
  &=& \frac{1}{\epsilon^2}\frac{1}{\xi^2+\eta^2}
     \biggl[ (1-\epsilon^2(\xi^2+1))(\xi^2+1)p_{\xi}^2 \biggl.\nonumber\\
  &+&\biggl. (1-\epsilon^2(1-\eta^2))(1-\eta^2)p_{\eta}^2 
   +     ( \frac{1}{1-\eta^2}-\frac{1}{\xi^2+1} ) 
         p_{\phi}^2  \biggl] \nonumber\\
  &=& \frac{\sigma^2}{\epsilon^2}\left[
        p_{x}^2+p_{y}^2+(1-\epsilon^2)p_{z}^2\right]
      - (x p_{x} + y p_{y} + z p_{z})^2
   \,.
   \label{bewegungskonstante.ellipt}
\end{eqnarray}
in elliptical and cartesian coordinates respectively.
This is the third independent 
constant of motion in addition to the energy $E$ and $L_{z}$.
This can be checked directly, using the equations of motion
for the time derivatives of $B\,$.
Similarly to the isotropic case conservation of $E$ and $B$ require
that trajectories hit the boundary orthogonally,
because the momenta there diverge.
In the isotropic limit $\sigma \to 0$ the elliptical 
coordinates become singular, 
and therefore it is more instructive to see this limit 
in cartesian coordinates. In this limit 
$\sigma /\epsilon$ is the Thomas-Fermi radius, and $B$ has
the simple meaning
\begin{equation}
   B = {2 E^2 \over \omega_0^2}+ L^2 \,.
    \label{beta.isotrop}
\end{equation}
%i.e., it is connected to the generalization of the total
%angular momentum square $L^2$ for the non-isotropic case.

The existence of the three independent constants 
of motion $E$, $L_z$ and $B$ explains the integrable
motion generated by $H_{hyd}$ and therefore the almost integrable
situation found numerically in the motion generated by the total Hamiltonian 
(\ref{bogol.ham.anders}) in the small energy and small angular momentum
region.
We notice that two kinds of trajectories can occur in this regime.
From (\ref{hamjacobi.ellipt}) we can determine the turning points
in $\xi$ and $\eta$.
In $\xi$-direction all the trajectories reach the Thomas-Fermi surface
and are reflected back there.
If the condition
\begin{equation}
   B > B^{*} = {2 E^2 \over \omega_0^2}+ L_z^2 \,,
    \label{beta.bifurc}
\end{equation}
is satisfied, there is an inner turning point in $\xi$-direction
and $\eta$ takes a range 
$[-\eta_{\mbox{\tiny max}},\eta_{\mbox{\tiny max}}]$.
These trajectories correspond to the hypocycloids
of the isotropic case,
as an example see Fig.\ref{hydro.trajects}a.
For $B<B^*$ however there are further turning points in $\eta$-direction,
the motion being confined between two hyperbolas with $\xi$ values
extending to zero,
which can be seen in Fig.\ref{hydro.trajects}b.
This kind of trajectory only occurs in the anisotropic system.
$B=B^*$ is the separatrix between these two types of motion.
As usual this separatrix is structurally unstable 
against small nonintegrable perturbations of the 
integrable motion in the hydrodynamic limit.
It plays a crucial role for the appearance of 
chaos in the Bogoliubov Hamiltonian as the energy is 
increased from values very small compared to $\mu$,
because it is destroyed and replaced by a chaotic 
separatrix layer, which is very narrow at first, 
but grows in width as the energy is increased.
In Fig.\ref{poinc.cuts}c 
two regular islands corresponding to the two 
kinds of trajectories and the chaotic separatrix layer between them  
can be seen.

%%%%%%%%%%%%%%%%%%%%%%%%%%%%%%%%%%%%%%%%
\subsection{Completely anisotropic case}

The analysis of the preceding section can be generalized to the case of
a completely anisotropic harmonic trap. The formulas become rather lengthy
and we just indicate the essential steps.

The trap potential is written in the form
\begin{equation}
\label{eq:44}
 U(\bbox{x})=\mu
  \left(\frac{x^2}{a^2}+\frac{y^2}{b^2}+\frac{z^2}{c^2}\right)
\end{equation}
with
\begin{equation}
\label{eq:45}
 a^2=2\mu/m\omega_x^2\quad,\quad (\mbox{\rm and cyclic})\,.
\end{equation}
We may assume $a>b>c$ without restriction of generality. Then new
elliptic coordinates $\xi$, $\eta$, $\zeta$ are introduced via
\begin{equation}
\label{eq:46}
 x=\pm
  \sqrt{\frac{(a^2+\xi)(a^2+\eta)(a^2+\zeta)}{(a^2-b^2)(a^2-c^2)}}
   \quad,\quad (\mbox{\rm and cyclic})
\end{equation}
after which the potential reads
\begin{equation}
\label{eq:47}
 U(\xi,\eta,\zeta)=\mu(1+\frac{\xi\eta\zeta}{a^2b^2c^2})\,.
\end{equation}
The range of $\xi,\eta,\zeta$ is $0\ge\xi\ge-c^2\ge\eta\ge-b^2\ge\xi\ge-a^2$.
The old canonical momenta $p_x$, $p_y$, $p_z$ are given in terms of the
new ones by
\begin{eqnarray}
 \label{eq:48}
  p_x=\sqrt{\frac{(a^2+\xi)(a^2+\eta)(a^2+\zeta)}{(a^2-b^2)(a^2-c^2)}}
  &&\Bigg[ 2p_\xi\frac{(b^2+\xi)(c^2+\xi)}{(\xi-\eta)(\xi-\zeta)}\nonumber\\
  && +2p_\eta\frac{(b^2+\eta)(c^2+\eta)}{(\eta-\zeta)(\eta-\xi)}\\
  && +2p_\zeta\frac{(b^É+\zeta)(c^2+\zeta)}{(\zeta-\xi)(\zeta-\eta)}
   \Bigg]\nonumber\\
  (\mbox{\rm and cyclic}) && \nonumber \,.
\end{eqnarray}
Then the Hamiltonian in the hydrodynamic limit can be written in terms
of the new canonical coordinates and momenta. 
The Hamilton-Jacobi equation
can be written in the nearly separated form
\begin{equation}
\label{eq:49}
 0=(\eta-\zeta)
  \left[(a^2+\xi)(b^2+\xi)(c^2+\xi)(\frac{\partial S}{\partial \xi})^2 +
   \frac{E^2m}{4\mu}\,\frac{a^2b^2c^2}{\xi}\right]+
    \,(\mbox{\rm cyclic})\,.
\end{equation} 
In fact complete separation is achieved, because this equation is satisfied
only by putting the angular brackets equal to $A+B\xi$ (and cyclic),
\begin{eqnarray}
\label{eq:50}
 &&\left[(a^2+\xi)(b^2+\xi)(c^2+\xi)p^2_\xi+
   \frac{E^2m}{4\mu}\,\frac{a^2b^2c^2}{\xi}\right]=\frac{1}{4}(A+B\xi)\\
  &&\qquad (\mbox{\rm and cyclic})\nonumber
\end{eqnarray}
where $A$ and $B$ are two separation constants, which are the same for all
three equations related by cyclic permutation. From these three equations
$A$ and $B$ can be eliminated by multiplying the first with $(\eta-\zeta)$,
the second with $(\zeta-\xi)$, the third with $(\xi-\eta)$ and adding
them. This gives, of course, back eq.~(\ref{eq:49}), which defines
$E=H_{\mbox{\tiny hyd}}$ in terms of the canonical variables. However,
solving the three equations instead for $A$ by eliminating $B$ and $E$,
and then for $B$ eliminating $A$ and $E$ we obtain two new conserved
phase-space functions. Translated back to Cartesian coordinates these read
\begin{eqnarray}
\label{eq:51}
A= && -
\left\{[(b^2+c^2)(x^2-a^2)+a^2(y^2+z^2)]p^2_x+2a^2yzp_yp_z
\right\}
+\,\mbox{\rm (cyclic)}
\end{eqnarray}
and
\begin{equation}
\label{eq:52}
  B = -\left\{(x^2-a^2)p^2_x+2yzp_yp_z\right\}+
   \,\mbox{\rm (cyclic)}\,.
\end{equation}
The conserved function $B$ is a simple generalization of the conservation
law we already found in the case with cylindrical symmetry, whereas $A$
corresponds to $L^2_z\,$.
By a straightforward but lengthy calculation it can be checked 
that the Poisson brackets
$\{H_{\mbox{\tiny hyd}}, A\}$, $\{H_{\mbox{\tiny hyd}}, B\}$, $\{A,B\}$
all vanish. Therefore, the dynamics governed by $H_{\mbox{\tiny hyd}}$
is still completely integrable even in this completely anisotropic case.

%%%%%%%%%%%%%%%%%%%%%%%%%%%%%%%%%%%%%%%%%%%%%%%%%%%%%%%%%%%%
\section{the Hartree-Fock dynamics}
%%%%%%%%%%%%%%%%%%%%%%%%%%%%%%%%%%%%%%%%%%%%%%%%%%%%%%%%%%%%

Another limiting case of the Bogoliubov description 
of quasiparticles (\ref{eq:9})
consists in neglecting the hole-component $V_j(\bbox{x})$ 
in the field operator $\hat{\varphi}(\bbox{x})$.
The remaining component $U_j(\bbox{x})$ is then described
by the Hartree-Fock Hamiltonian (\ref{eq:10}).
The interaction between particles is taken into account
by the potential $K(\bbox{x})$,
describing
the mean interaction of one particle
with all the other particles.
Restricting ourselves to $T=0$ all those other particles 
are in the condensate.
In the homogenous systems this approach simply results
in a shift of the dispersion relation of noninteracting particles by the chemical potential $\mu$.
For spatially homogeneous Bose condensates and also Bose condensates in traps such a description can be applied for energies 
larger than the mean interaction energy given by $\mu$. 
However, in traps there is even a regime for energies smaller than $\mu$ where the Hartree-Fock approximation applies \cite{dalfovo}, 
namely in the case when the kinetic energy $\epsilon_{kin}(\bbox{p})$ 
is large compared to the {\it local} mean interaction energy $K(\bbox{x})$. This can be satisfied in a layer around the surface of the Bose condensate where $K(\bbox{x})$ is very small.

Using the Thomas-Fermi approximation for the wave function
(\ref{eq:5})
the Bogoliubov Hamiltonian and the Hartree-Fock Hamiltonian coincide outside the condensate.
Inside the condensate, 
if the kinetic energy $\epsilon_{kin}(\bbox{p})$ 
is much larger than the potential term  $K(\bbox{x})$,
an expansion of the Bogoliubov Hamiltonian
(\ref{bogol.ham.anders}) to first order in $K(\bbox{x})$ just gives
the Hartree-Fock Hamiltonian 
\begin{equation}
   H_{\mbox{\tiny HF}}=\frac{p^2}{2m}+|U(\bbox{x})-\mu|
   \,,
   \label{hartree.ham}
\end{equation}
which is therefore valid, for $\epsilon_{kin}>>K(\bbox{x})$, inside and outside the condensate.

We now want to investigate the classical dynamics 
of this Hartree-Fock Hamiltonian.
The isotropic problem is completely integrable again.
As constants of motion we can take the energy, 
the modulus and the $z$-component of the angular momentum.
We immediately turn to the classical dynamics 
of the anisotropic, but axially symmetric case
in the trap potential (\ref{eq:24})
and consider it as
a system with two degrees of freedom.
The conserved angular momentum around the symmetry axes $L_z$
enters only as a parameter.
Again we investigate the dynamics by Poincar\'e cuts, now taken  
on the Thomas-Fermi surface $\xi=(1/\epsilon^2-1)^{1/2}$ and parametrized by the second elliptical coordinate $\eta$ 
and its conjugate momentum $p_{\eta}$.
For energies much larger than the chemical potential
the interaction with the condensate is only a small perturbation
to the integrable motion in the harmonic trap
and we observe quasi-regular behaviour.
In this limit the Bogoliubov description of quasi-particles
reduces to the Hartree-Fock description,
the condition $\epsilon_{kin}(\bbox{p}) \gg K(\bbox{x})$
being fulfilled for all trajectories,
and the classical motions generated by both Hamiltonians are 
essentially the same.
Trajectories not entering the condensate are even identical,
since here the two descriptions fully coincide.

For energies in the approximate range $10>E/\mu>0.1$
we observe a mixed phase space again (see Fig.\ref{poinc.hartree}a).
A regular island around the periodic orbit $z=0=p_z$
is surrounded by a chaotic sea.
For $E>\mu$ the structure in phase space 
is similar as for the Bogoliubov dynamics,
but differing in detail.
From this we can conclude that the stochasticity
observed for the Bogoliubov Hamiltonian in Fig.\ref{poinc.cuts}
is not a consequence of the coupled two component structure
of the underlying semiclassical description,
but is simply caused by the anisotropy of the external potential.

For energies much smaller than the chemical potential 
$E<0.1\mu$
we find regular behaviour again,
see Fig.\ref{poinc.hartree}b.
Particles are confined to the sharp potential channel
near the Thomas-Fermi surface.
The width of this potential channel scales as $E/\mu$.
Roughly the particles spend the same time outside and inside the condensate.
We look at the problem in elliptical coordinates (\ref{ellipt.coords}),
and choose for concreteness again the case $\omega_{z} > \omega_0$.
The oscillations in $\xi$ orthogonal to the Thomas-Fermi surface
$\xi=(1/\epsilon^2-1)^{1/2}$ are
much faster than the oscillations in $\eta$-direction 
along the channel.
This suggests to make an adiabatic approximation in which the action-integral $I_{\xi}=(2\pi)^{-1}\oint p_{\xi}d\xi$ over one full cycle in $\xi$ at fixed $\eta$, $p_{\eta}$ emerges
as an adiabatic constant for the motion. Evaluating this adiabatic invariant
for $E/\mu \ll 1$ we get as a function of $\eta$, $p_{\eta}$
\begin{equation}
\label{ixi}
I_{\xi}=\frac{4\mu}{3\pi\omega_z}\,
\frac{1}{\sqrt{1-\epsilon^2(1-\eta^2)}}\,
\Bigl(\frac{E}{\mu}-\frac{1-\eta^2}{1-\epsilon^2(1-\eta^2)}
(\frac{\omega_0p_{\eta}}{2\mu})^2-\frac{1}{1-\eta^2}
(\frac{\omega_0 L_{z}}{2\mu})^2\Bigr)^{3/2}
\,.
\end{equation} 
This new adiabatically conserved quantity which emerges 
in the low-energy limit of the Hartree-Fock dynamics, 
is the cause of integrability in that limit.

Solving this equation for the energy we get the Hamiltonian
of the slow $\eta$-dynamics, valid for low energies.
From (\ref{ixi}) we see that the turning points in $\eta$
are independent of the energy if we keep $L_z/E^2$
and $I_{\xi}/E^{3/2}$ constant.
In Fig.\ref{poinc.hartree}b we compare a Poincar\'e section 
in $\eta$ and $p_{\eta}$
of the dynamics of the Hartree-Fock Hamiltonian
with trajectories of the slow $\eta$-dynamics
for different values of $I_{\xi}$.
Both curves agree very well.
For smaller energies $E \le 0.01\mu$
no difference between both curves can be noticed.

Now we have to ask ourselves, which of the trajectories displayed in Fig.\ref{poinc.hartree}b are indeed good approximations to trajectories described by the Bogoliubov Hamiltonian. Let us look first at the isotropic case,
where the motion separates in radial and angular motion.
The kinetic energy in the angular degree of freedom roughly is
$L^2/2mr^2 \approx (\omega_0 L)^2/4\mu$.
Since for low energies $r$ deviates only very little
from the Thomas-Fermi radius
this rotational energy is almost conserved.
The remaining energy is stored
in the radial degree of freedom and
only this energy can be transformed to potential energy.
So the condition 
that the Hartree-Fock dynamics 
and the Bogoliubov dynamics agree is in this case
\begin{equation}
\label{condition}
E-\frac{(\omega_0 L)^2}{4\mu} \ll E
\,.
\end{equation}  
For the anistropic case we can formulate an analogous criterion:
Only if most of the energy is kinetic energy of the motion 
parallel to the boundary, which cannot be transformed 
into potential energy,
the approximation of the Bogoliubov dynamics 
by the Hartree-Fock dynamics works well.
For $I_{\xi}=0$ no motion takes place orthogonal to the surface,
$\xi$ being constant,
and all the energy is stored in motion parallel to the surface.
This corresponds to the outer orbit forming the boundary of the cut 
in Fig.\ref{poinc.hartree}.
The maximal value of $I_{\xi}$ for fixed $L_z$ 
is given by setting
$\eta = p_{\eta} = 0$ in (\ref{ixi}).
This trajectory corresponds to the origin 
of Fig.\ref{poinc.hartree},
where motion takes place only in $\xi$ and $\phi$-directions.
Generally the two dynamics differ in this case,
unless most of the energy is stored 
in angular motion of the cyclic variable $\phi$ which is 
also motion along the Thomas-Fermi surface.
The maximal value of $I_{\xi}$ for a given energy
is found by neglecting both angular motions in $\eta$
and in $\phi$ in (\ref{ixi}), by putting $p_{\eta},L_z$ 
equal to zero there.
As a condition that only a small fraction of energy is 
stored in the motion
orthogonal to the surface
and hence that
both motions from the Bogoliubov and from the Hartree-Fock 
Hamiltonian agree,
we can thus state:
\begin{equation}
\label{condition.aniso}
I_{\xi} \ll I_{\xi}^{max} = \frac{4\mu}{3\pi\omega_0}\,
\Bigl(\frac{E}{\mu}\Bigr)^{3/2}
\,.
\end{equation} 
In Fig.\ref{poinc.compare} we compare Poincar\'e cuts
of the Bogoliubov dynamics with the one-degree of freedom motion
obtained from $(\ref{ixi})$, representing 
the integrable Hartree-Fock dynamics for small energies.
We see that indeed both dynamics agree well for small values
of $I_{\xi}$ near the boundary of the cut.
For $L_z$ chosen large, see Fig.\ref{poinc.compare}a,
even for values of $I_{\xi}$ close to the maximal one,
both dynamics in $\eta$ agree qualitatively.
However the different behaviour in the variable $\xi$
orthogonal to the surface can, of course,
not be seen in this cut at constant $\xi$.
For smaller values of $L_z$, see Fig.\ref{poinc.compare}b,
we can distinguish two regions.
Near the boundary, for small $I_{\xi}$,
we see the Hartree-Fock limit of the Bogoliubov dynamics,
where both dynamics agree.
The inner region corresponds to the hydrodynamic limit
of the Bogoliubov dynamics
and cannot be compared with the Hartree-Fock dynamics.
The two kinds of closed tori visible here are the two kinds
of hydrodynamic trajectories discussed at the end of section III.C..

%%%%%%%%%%%%%%%%%%%%%%%%%%%%%%%%%%%%%%%%%%%%%%%%%%%%%%%%%
\section{conclusions}

The quasi-particle excitations are the basic constituents of 
the dynamical and thermodynamical properties of Bose condensates. 
In the present paper we have investigated their dynamics for Bose 
condensates of atomic gases in traps in the classical limit. 
The two limiting types of excitations, collective modes and 
quasi-particle excitations consisting essentially of single atoms 
moving in a mean field correspond, in the classical limit, to 
particles and anti-particles of zero mass, moving 'relativistically' 
with the speed of sound, and to single atoms moving in the potential 
created by the trap and the Hartree-Fock potential energy of all other 
atoms. In spatially homogeneous (untrapped) condensates these two 
types of excitation strongly differ in energy $E$, the collective 
modes occurring at $E\ll\mu$, the single-particle modes at $E\gg\mu$. 
In the trapped condensates both types of excitations coexist, at least 
classically, at small energies $E\ll\mu$, and are instead spatially 
separated. The collective modes
live inside the condensate, the single-particle modes at small 
energies in a narrow layer at the border.

One principal result we have obtained here is that the classical 
dynamics of both, the collective modes and the single-particle modes, 
become integrable in the limit $E/\mu\ll1$. This has important 
consequences for the quantum dynamics as well: the integrability 
can be used there to separate the 
Schr\"odinger equation and to obtain not only the low-lying levels 
of the collective modes \cite{ohberg,fliesser}, but also of the 
single-particle modes. After quantization an energy gap reappears 
separating the collective modes with typical energies $\hbar\omega_0$ 
and the single-particle modes whose lowest levels have energies of the 
order $(\hbar\omega_0)^{2/3}\mu^{1/3}$ due to their close confinement
in normal direction to the surface of the condensate. However, this 
energy difference is much smaller than, and has a different origin 
as the energy difference between both types of modes in homogeneous systems.

Another principal result obtained here is the {\it nonintegrabilty} of the 
classical dynamics of the quasi-particle excitations at intermediate 
energies $E\simeq\mu$. This applies to both, the full Bogoliubov 
dynamics and the limiting Hartree-Fock dynamics approximating it 
wherever the kinetic energy is large compared to the 
{\it local} mean interaction energy. Again this nonintegrability 
has a direct consequence also for the quantum dynamics, because it 
implies avoided crossings between quasi-particle levels as functions 
of the dimensionless interaction strength $N_0a/d_0$ with 
$d_0=\sqrt{\hbar/m\omega_0}$, if the energy and $\mu$
are comparable. Such avoided crossings have indeed been 
seen in numerically generated plots\cite{you}.

Our results not only explain these avoided crossings, 
they also open the door to an intreaguing wider perspective, 
quantum chaos of the quasi-particle dynamics
in the Bose condensates of atoms in anisotropic traps. 
%%%%%%%%%%%%%%%%%%%%%%%%%%%
\section*{Acknowledgements}

This work has  
been supported by the
project
of the Hungarian Academy of Sciences (grant No. 95) and the Deutsche
Forschungsgemeinschaft.
M.F.\ gratefully acknowledges support
by the German-Hungarian Scientific and Technological Cooperation
under Project 62.
R.G. and M. F. wishes to acknowledge support by the Deutsche
Forschungsgemeinschaft through the Sonderforschungsbereich 237
``Unordnung und gro{\ss}e Fluktuationen''.
Two of us (A.Cs,P.Sz) would like to acknowledge support by
the Hungarian Academy of Sciences under grant No. AKP 96-12/12.
The work has been partially supported by the Hungarian
National Scientific Research Foundation under grant
Nos. OTKA T017493, F020094 and by the Ministry of Education of
Hungary under grant No. FKFP0159/1997.

\begin{figure}
\caption{Poincar\'e sections of the dynamics 
of the Bogoliubov Hamiltonian
(\protect\ref{eq:20}) in cylindrical coordinates
for the different energies (from top to bottom) $E/\mu=40$ (a), 
1 (b) and 0.02 (c).
The cut is taken at $z=0$ and diplayed in the variables
$\rho,p_{\rho}$ in units of $(2\mu/m\omega_0^2)^{1/2}$, $(2m\mu)^{1/2}$, respectively.
The anisotropy is chosen as $\omega_z/\omega_0 = \protect\sqrt{8}$,
the angular momentum was fixed as $\omega_0L_z/E=0.2$.
\label{poinc.cuts}}
\end{figure}

\begin{figure}
\caption{Trajectories in coordinate space 
of the Bogoliubov dynamics of (\protect\ref{eq:20})
starting from the same point and in the same direction
for different energies $E/\mu=0.1$ (dashed line), 
0.01 (solid) and $10^{-6}$(dotted).
\label{limit.trajects}}
\end{figure}

\begin{figure}
\caption{Coordinates for the hypocycloid (\protect\ref{hypocikloid})
\label{pic.cycloid}}
\end{figure}

\begin{figure}
\caption{Trajectory 
of the hydrodynamic Hamiltonian (\protect\ref{hydro.ham})
for $B>B^*$(a) and for $B<B^*$(b). $z,\rho$ are plotted in units of $(2\mu/m\omega_0^2)^{1/2}$.
\label{hydro.trajects}}
\end{figure}

\begin{figure}
\caption{Poincar\'e section of the dynamics 
of the Hartree-Fock Hamiltonian
(\protect\ref{hartree.ham}) in elliptical coordinates
(\protect\ref{ellipt.coords})
for the energy $E/\mu=1$(a) and $E/\mu=.06$(b).
The cut is taken on the Thomas-Fermi surface
in the variables $\eta,p_{\eta}$
for $\omega_z/\omega_0 = \protect\sqrt{8}$.
The angular momentum is given by $(\omega_0 L_z)^2/2\mu E=1$.
Solid lines in (b) are trajectories of the Hamiltonian
in $(\eta,p_{\eta})$
following from (\protect\ref{ixi}).
\label{poinc.hartree}}
\end{figure}

\begin{figure}
\caption{Poincar\'e sections of the dynamics 
of the Bogoliubov Hamiltonian
(\protect\ref{eq:20}) in elliptical coordinates
(\protect\ref{ellipt.coords})
for the energies $E/\mu=0.1$ with $(\omega_0L_z)^2/2\mu E=1$(a)
and $E/\mu=0.01$ with $(\omega_0L_z)^2/2\mu E=.02$(b)
on the Thomas-Fermi surface
in $\eta,p_{\eta}$
for $\omega_z/\omega_0 = \protect\sqrt{8}$.
Solid lines are trajectories of the Hamiltonian
in $(\eta,p_{\eta})$
following from (\protect\ref{ixi}).
\label{poinc.compare}}
\end{figure}


\begin{references}
\bibitem{anderson} M.~H.~Anderson, J.~R.~Ensher, M.~R.~Matthews,
                   C.~E.~Wiemann, and E.~A.~Cornell, 
                   Science {\bf 269}, 198 (1995).
\bibitem{bradley} C.C.~Bradley, C.A.~Sackett, J.J.~Tollett, and R.G.~Hulet,
                  Phys.\ Rev.\ Lett.\ {\bf 75}, 1687 (1995).
\bibitem{davis} K.B.~Davis, M.O.~Mewes, M.R.~Andrews, N.J.~van Druten,
                D.D.~Durfee, D.M.~Kurn, and W.~Ketterle,
                Phys.\ Rev.\ Lett.\ {\bf 75}, 3969 (1995).

\bibitem{gross} L.~P.~Pitaevskii, Zh.~Eksp.~Teor.~Fiz.~{\bf 40}, 646 (1961)
                [Sov.~Phys.~JETP~{\bf 13}, 451 (1961)];
                E.~P.~Gross, Nuovo Cimento~{\bf 20}, 454 (1961);
                J.~Math.~Phys.~{\bf 4}, 195 (1963).
\bibitem{baym}G.~Baym and C.~J.~Pethick,
              {\it Phys.~Rev.~Lett. \bf 76}, 6 (1996)
\bibitem{fetter} A.~L.~Fetter, Ann.~Phys.~(N.Y.)~{\bf 70}, 67 (1972).

\bibitem{numerical} M.~Edwards, 
                    P.~A.~Ruprecht and K.~Burnett,
                    R.~J.~Dodd and C.~W.~Clark,
                    {\it Phys.~Rev.~Lett.} {\bf 77}, 1671 (1996); 
                    P.~A.~Ruprecht, M.~Edwards, and K.~Burnett,
                    Phys.~Rev.~{\bf A 54}, 4178 (1996);
                    K.~G.~Singh and D.~S.~Rokhsar,
                    {\it Phys.~Rev.~Lett.} {\bf 77}, 1667 (1996); 
                     L.~You, W.~Hoston, and M.~Lewenstein,
                    {\it Phys.~Rev. \bf A55}, R1581 (1997);
                    A.~Smerzi and S.~Fantoni,
                    {\it Phys.~Rev.~Lett.} {\bf 78}, 3589 (1997);
                    B.~D.~Esry,
                    {\it Phys.~Rev. \bf A55}, 1147 (1997).

\bibitem{analytical} A.~L.~Fetter,
                     {\it Phys.~Rev. \bf A53}, 4245 (1996);
                     W.-C.~Wu and A.~Griffin,
                     {\it Phys.~Rev. \bf A54}, 4204 (1996);
                     V.~M.~P\'erez-Garc\'\i a, H.~Michinel,J.~I.~Cirac,
                      M.~Lewenstein, P.~Zoller,
                     {\it Phys.~Rev.~Lett \bf 77}, 1520 (1996);
                     Y.~Castin and R.~Dum,
                     {\it Phys.~Rev.~Lett. \bf 77}, 5315 (1996);
                     Yu.~Kagan, E.~L.~Surkov, G.~V.~Shlyapnikov,
                     {\it Phys.~Rev. \bf A54}, R1753 (1996);
                     {\it Phys.~Rev. \bf A55}, R18 (1997);
                     F.~Dalfovo, C.~Minniti, 
                     S.~Stringari, L.~P.~Pitaevskii,
                     Phys.~Lett.~{\bf A227} 259 (1997).

\bibitem{csordas} A.~Csord\'as, R.~Graham, P.~Sz\'epfalusy,
                  cond-mat9705133 (unpublished).
\bibitem{stringari} S.~Stringari, Phys.\ Rev.\ Lett.\ {\bf 77}, 2360 (1996).
\bibitem{dalfovo} F.~Dalfovo, S.~Giorgini, L.~P.~Pitaevskii and S.~Stringari,
                  cond-mat/9705240 (unpublished).
\bibitem{ohberg} P.~\"Ohberg, E.~L.~Surkov, I.~Tittonen,
                 M.~Wilkens, and G.~V.~Shlyapnikov, 
                 physics/9705006 (unpublished).
\bibitem{fliesser} M.~Fliesser, A.~Csord\'as, R.~Graham, P.~Sz\'epfalusy,
                  cond-mat/9706002 (unpublished).
\bibitem{you} L.~You, W.~Hoston, M.~Lewenstein, and M.~Marinescu, 
              preprint (1997).

\end{references}
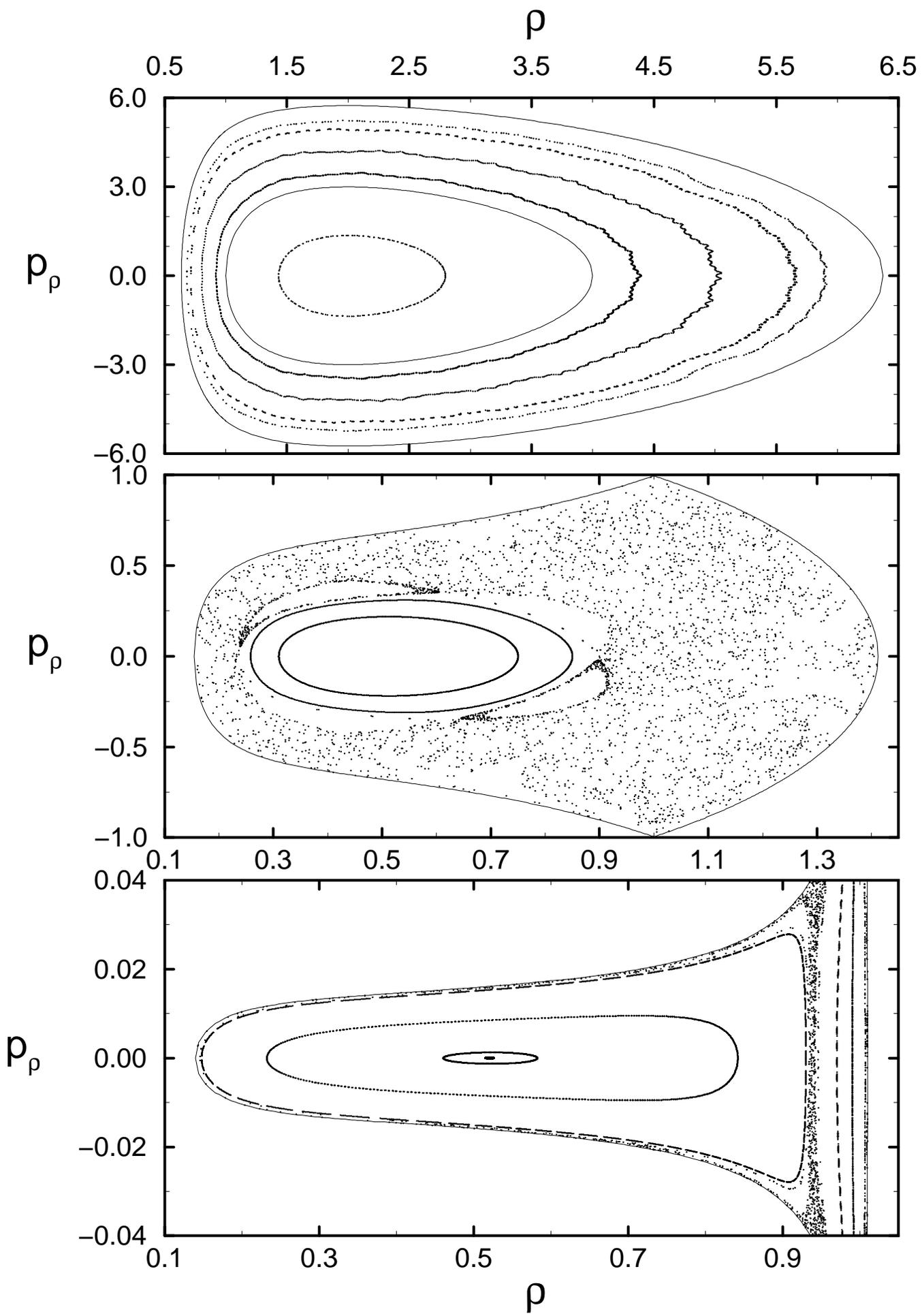
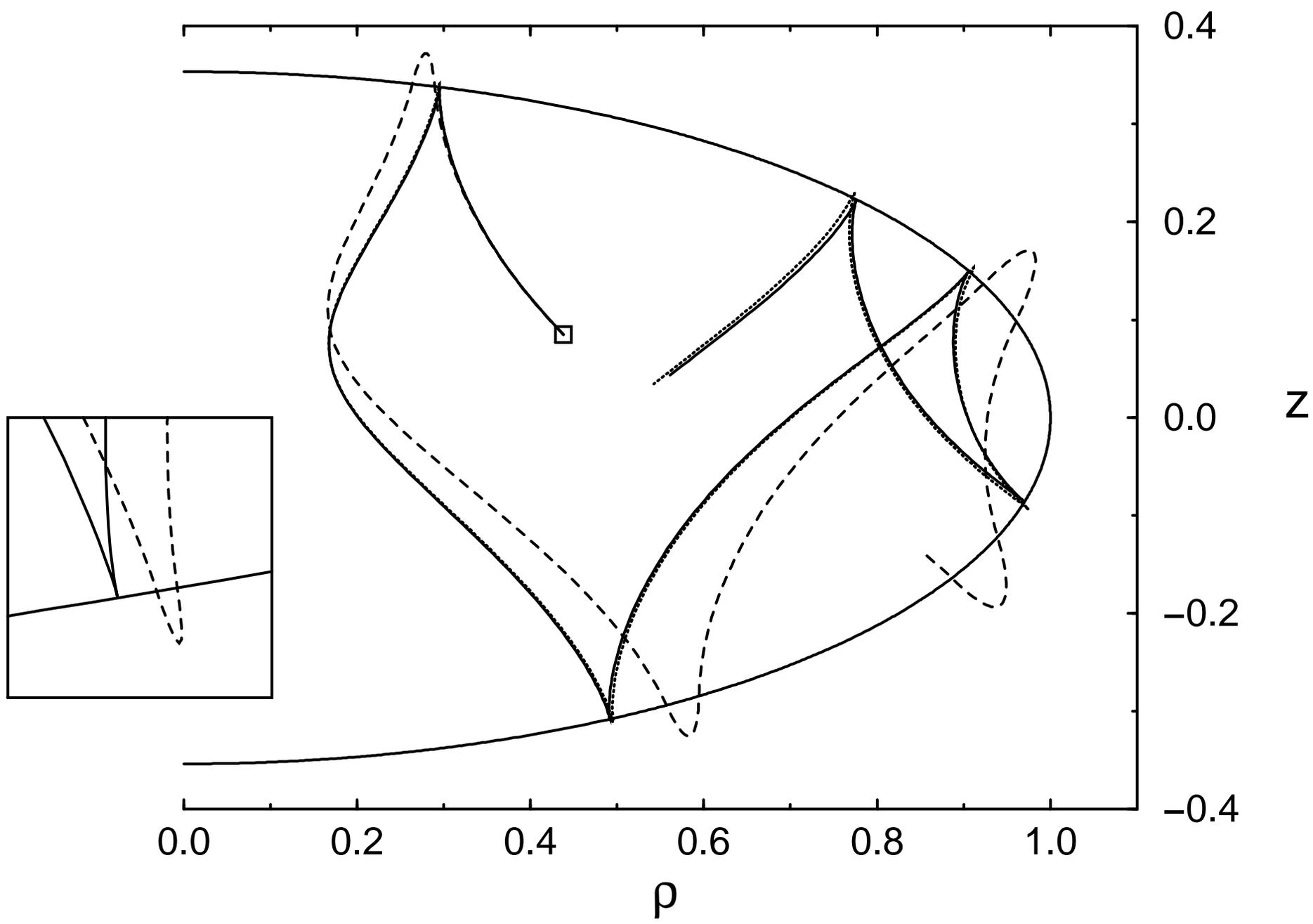
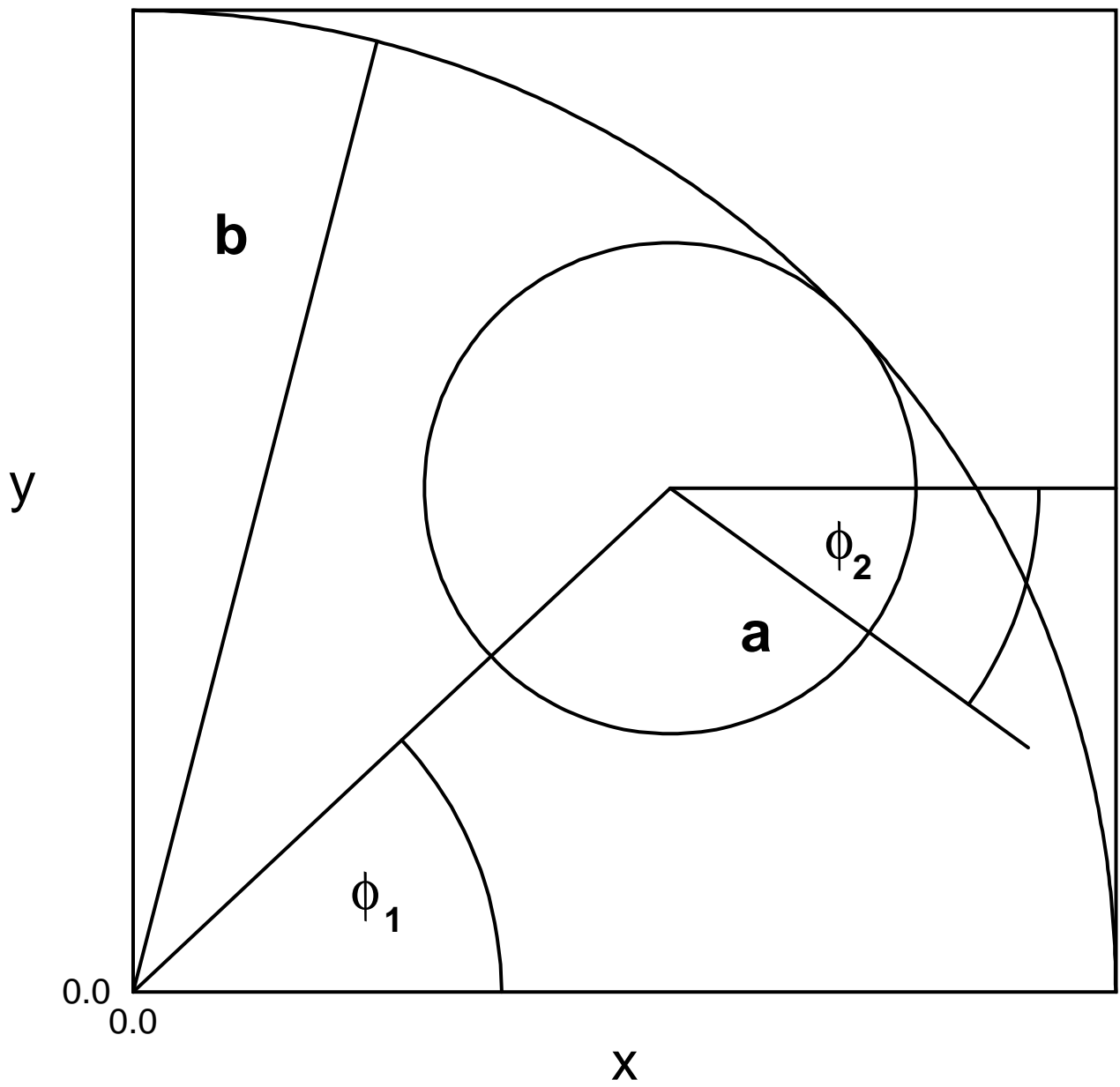
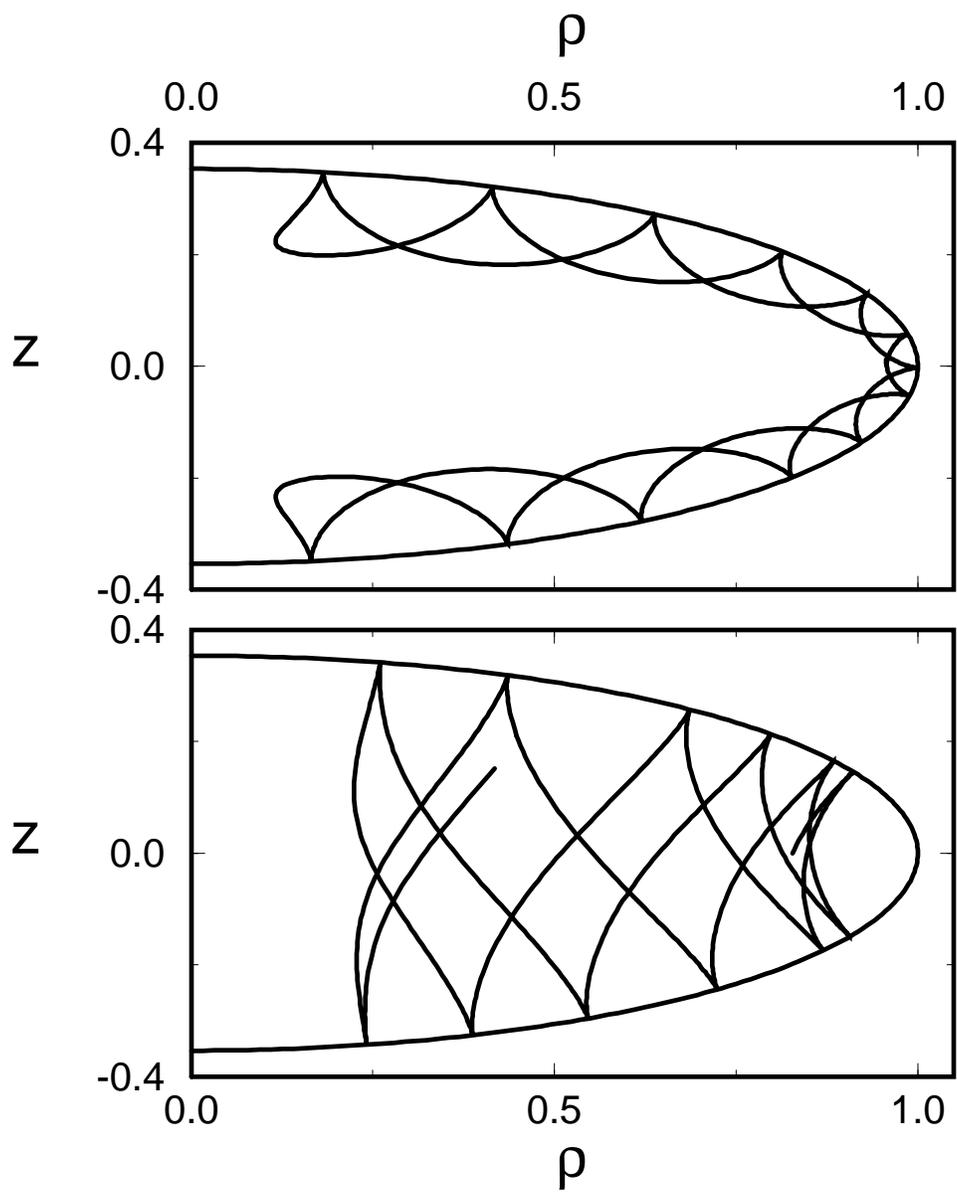
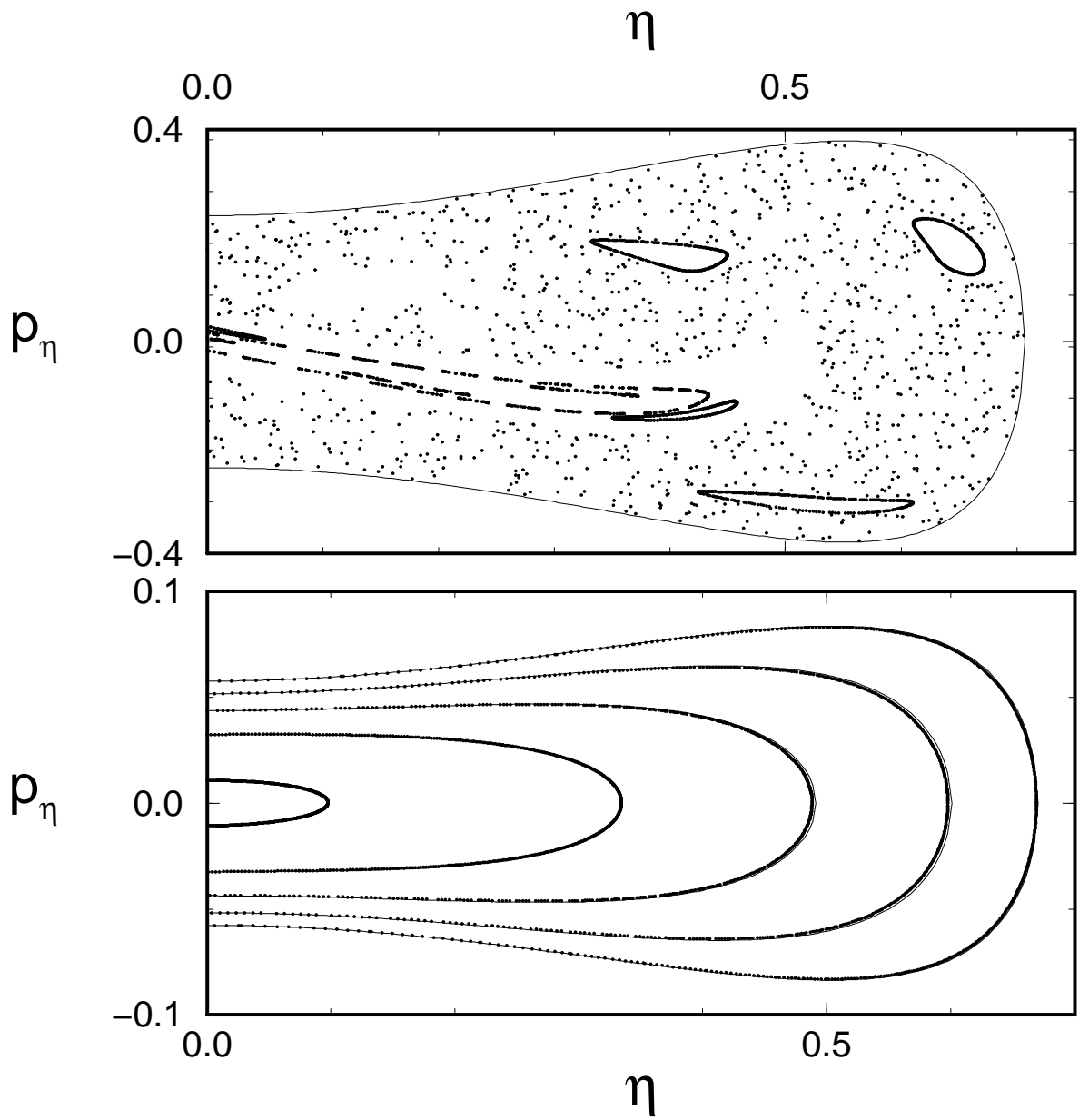
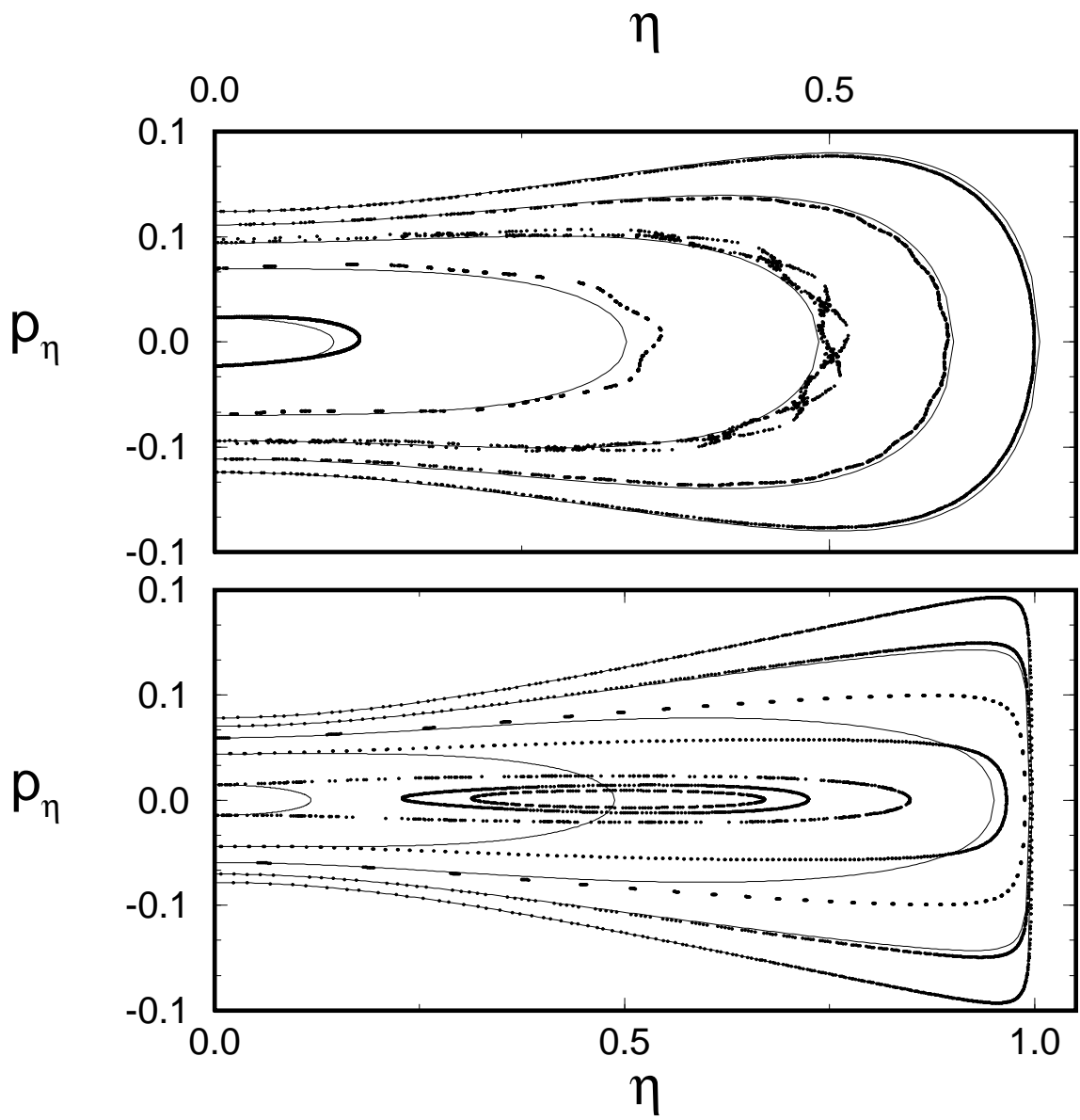
\end{document}